\documentstyle[12pt,a4,epsfig,a4wide]{article} 
\def\Jo#1#2#3#4{{#1} {\bf #2}, #3 (#4)}

\def\NPB{{Nucl. Phys.} {\bf B}}
\def\NPBP{{Nucl. Phys.} {\bf B} (Proc. Suppl.)}

\def\PLB{{Phys. Lett.}  {\bf B}}
\def\PRL{Phys. Rev. Lett.}
\def\PRD{{Phys. Rev.} {\bf D}}

\def\PRP{{Phys. Rep. }}

\def\ZPC{{Z. Phys.} {\bf C}}

\def\PNPP{Prog. Nucl. Part. Phys.}

\def\IJMP{Int. J. Mod. Phys. {\bf A}}

\def\ra{\rightarrow}
\def\be{\begin{equation}}
\def\ee{\end{equation}}
\def\gs{\mathrel{
   \rlap{\raise 0.511ex \hbox{$>$}}{\lower 0.511ex \hbox{$\sim$}}}}
\def\ls{\mathrel{
   \rlap{\raise 0.511ex \hbox{$<$}}{\lower 0.511ex \hbox{$\sim$}}}}

\newcommand{\onbb}{neutrinoless double beta decay}

\newcommand{\ba}{\begin{array}{c}}
\newcommand{\baz}{\begin{array}{cc}}
\newcommand{\bad}{\begin{array}{ccc}}
\newcommand{\bav}{\begin{array}{cccc}}
\newcommand{\bea}{\begin{equation} \begin{array}{c}}
\newcommand{\eea}{ \end{array} \end{equation}}
\newcommand{\ea}{\end{array}}
\newcommand{\D}{\displaystyle}

\newcommand{\dma}{\mbox{$\Delta m^2_A$}}
\newcommand{\dms}{\mbox{$\Delta m^2_\odot$}}

\begin{document}

\title{  
\hfill { \bf {\small hep-ph/0104228}}\\ 
\hfill { \bf {\small DO--TH 01/05}}\\ \vskip 1cm 
\bf Leptogenesis in left--right symmetric theories}
\author{Anjan S.~Joshipura$^a$\thanks{E--mail: \tt anjan@prl.ernet.in}, 
$\;$ Emmanuel A.~Paschos$^b$\thanks{E--mail: \tt paschos@physik.uni-dortmund.de}, $\;$ Werner Rodejohann$^b$\thanks{E--mail: \tt 
rodejoha@xena.physik.uni-dortmund.de}\\ \\
{ \normalsize \it $^a$Theoretical Physics Group,} 
{ \normalsize \it Physical Research Laboratory,}\\ 
{\normalsize \it Navarangapura, Ahmedabad, 380 009, India}\\ \\
{\normalsize \it $^b$Institut f\"ur Theoretische Physik, 
Universit\"at Dortmund,}\\
{\normalsize \it Otto--Hahn--Str.4, 44221 Dortmund, Germany}}
\date{}
\maketitle
\thispagestyle{empty}
\begin{abstract}

The masses and mixing of the light left--handed neutrinos
can be related to those of the heavy right--handed neutrinos 
in left--right symmetric theories. Properties of the
light neutrinos are 
measured in terrestrial 
experiments and the $CP$--violating decays of their heavy 
counterparts produce 
a baryon asymmetry via the well--known leptogenesis mechanism.  
The left--handed 
Higgs triplet, present in left--right symmetric theories, modifies the usual
see--saw formula. It is possible to relate
the lepton asymmetry to the light neutrino parameters when the triplet
and the top quark through the usual see--saw mechanism give the dominant
contribution to the neutrino mass matrix. We find that in 
this situation the small angle MSW 
and vacuum solutions produce reasonable asymmetry, whereas the large angle 
MSW case requires extreme fine--tuning of the three phases in the 
mixing matrix.

\end{abstract}
%Keywords:\\
%PACS:  

\newpage   

\section{\label{sec:intro}Introduction}

The explanation of the observed ratio of baryons to photons in the 
universe is 
one of the most challenging theoretical problems. 
In standard cosmology the ratio is 
explained as a disappearance of antimatter in the early universe as 
proposed by 
Sakharov \cite{sakharov}. The creation of a matter--antimatter 
asymmetry is, in many 
cases, suppressed by the conservation of the $B-L$ quantum number. 
Fukugita and Yanagida observed \cite{first} that a Majorana mass term 
provides an attractive 
possibility for the creation of a lepton asymmetry when heavy 
Majorana neutrinos 
decay at an epoch in which they are out of equilibrium. The 
effect is further enhanced by 
self--energy contributions which create relatively long--lived 
states \cite{leptogenesis}. 
The asymmetry is later converted into a baryon asymmetry via sphaleron 
processes \cite{sphaleron}. This Majorana neutrinos come closer 
to explaining the 
observed ratio of baryons to photons of \cite{citeYB} 
\be \label{eq:YBexp}
Y_B \simeq (0.1 \ldots 1) \cdot 10^{-10} . 
\ee

The explanation of the baryon asymmetry seems to demand physics beyond the 
Standard Model \cite{SMnot}. 
In addition, the collected evidence for massive neutrinos 
\cite{nurev} also demands physics beyond the Standard Model. The next 
logical step is to check if one can relate 
the data of the light left--handed neutrinos with the 
heavy right--handed ones and to obtain the correct 
order of magnitude for $Y_B$. 
Several recent papers dealt with this problem 
\cite{lola,laza1,others,kang,goldberg,jean,nezri,faltra1,faltra2,nielsen}, 
assuming specific structures for the mass matrices and 
symmetries of the theory. 
In this article we study a left--right symmetric model where in 
addition to the 
usual Higgs doublet there are left-- and right--handed Higgs triplets. 
The breaking of the 
symmetry generates vacuum expectation values $v_L$ and $v_R$ 
which in turn generate 
neutrino mass matrices. For a natural choice of parameters, 
the left--handed Higgs 
triplet gives the main contribution to the neutrino mass matrix. 
Only the top quark contribution of the Dirac mass matrix entering 
through the see--saw mechanism is of comparable size. 
The important role played by the triplet Higgs was 
highlighted in
\cite{eapjos}. 
In this case, the light 
left-- and heavy right--handed neutrino sector are related naturally and 
no further assumptions are required. At the end we find that from the three 
solutions to the solar 
neutrino problem small angle MSW and vacuum oscillations generate a 
baryon asymmetry of the correct order of magnitude. The large mixing 
angle MSW solution yields a very high $Y_B$.\\

The paper is organized as follows: In Section \ref{sec:convss} we review 
the conventional see--saw mechanism and its application to leptogenesis. In  
Section \ref{sec:lrmodels} we describe how the mechanism is modified 
in left--right symmetric 
theories. The experimental status of the 
left--handed 
neutrino mass matrix is included in Section \ref{sec:mnu}, 
which is then used to calculate the right--handed mass matrix 
in Section \ref{sec:detf}. These results are collected together in 
Section \ref{sec:res} and figures for the asymmetry as function of 
the parameters 
are presented. In the last Section \ref{sec:concl} we give our conclusions.

\section{\label{sec:convss}Conventional See--Saw mechanism}
The conventional see--saw mechanism follows from the Lagrangian of the 
Standard Model 
enlarged by the addition of a singlet right--handed  
neutrino $N'_{Ri}$ for each generation. The new part of the Lagrangian is 
\be \label{eq:convss}
-\mbox{$\cal{L}$}_Y 
= \overline{l_{iL}} \, \frac{\D \Phi}{\D \langle \Phi \rangle} \, 
\tilde{m}_{D ij} \, N'_{Rj}
+ \frac{\D 1}{\D 2}\overline{N_{Ri}^{c'}} \, M_{R ij} \, N'_{Rj} + \; \rm h.c.
\ee
with $l_{iL}$ the leptonic doublet, $\langle \Phi \rangle$ the 
vacuum expectation value (vev) of the 
conventional Higgs doublet $\Phi$, $\tilde{m}_D$ a Dirac mass matrix 
operating in generation space and $M_R$ is the symmetrical 
Majorana mass matrix for the right--handed neutrinos. 
We can go to the physical basis by diagonalizing $M_R$
\be
U_R^\ast \, M_R \, U_R^\dagger = {\rm diag}(M_1, M_2, M_3) 
\ee
and defining the physical states 
\be
N_R = U_R \, N'_R . 
\ee
In the new basis the Dirac mass matrix also changes to
\be \label{eq:mdur}
m_D = \tilde{m}_D \, U_R  . 
\ee
Thus, the Dirac Yukawa couplings are also rotated by the 
matrix $U_R$.\\

Interference of tree level with 
one--loop vertex and self--energy diagrams 
leads to a lepton asymmetry in 
the decays of the lightest Majorana, $N_1 \ra \Phi \, l^c$ and 
$N_1 \ra \Phi^\dagger \, l$ \cite{leptogenesis}:
\be \label{eq:eps}
\varepsilon = \frac{\D \Gamma (N_1 \ra \Phi \, l^c) - 
\Gamma (N_1 \ra \Phi^\dagger \, l)}{\D \Gamma (N_1 \ra \Phi \, l^c) +  
\Gamma (N_1 \ra \Phi^\dagger \, l)}
= \frac{\D 1}{\D 8 \, \pi \, v^2} \frac{\D 1}{(m_D^\dagger m_D)_{11}} 
\sum_{j=2,3} {\rm Im} (m_D^\dagger m_D)^2_{1j} f(M_j^2/M_1^2)
\ee
where $v\simeq 174$ GeV is the weak scale and 
the function $f$ is defined as 
\[
f(x) = \sqrt{x} \left(1 + \frac{1}{1 - x} - 
(1 + x) \, \ln \left(\frac{1 + x}{x}\right) \right)
\simeq -\frac{3}{2 \, \sqrt{x}} . 
\]
The approximation holds for $x \gg 1$.  
There can be a resonant enhancement of the asymmetry in case of the
degenerate Majorana neutrinos. 
Obviously, the magnitude of the asymmetry is of great interest since it 
introduces a $B-L$ violation in the theory.\\

As  already mentioned, the interaction in Eq.\ (\ref{eq:convss}) 
leads to the famous see--saw prediction for the light neutrino 
mass matrix \cite{seesaw} 
\be \label{eq:convmnu}
m_\nu = - \tilde{m}_D \, M_R^{-1} \, \tilde{m}_D^T = 
- m_D  \, {\rm diag}(M_1^{-1}, M_2^{-1}, M_3^{-1}) \, m_D^T . 
\ee
Note that $\tilde{m}_D$ in Eq.\ (\ref{eq:convss}) can always be written as
$\tilde{m}_D=V_L \, {\rm diag}(m_{1D},m_{2D},m_{3D}) \, V_R^{\dagger}$.
It then follows from Eqs.\ (\ref{eq:mdur},\ref{eq:eps}) that the asymmetry
$\varepsilon$ depends upon
the right--handed mixing matrices $V_R$ and $U_R$ rather than the
experimentally 
accessible left--handed ones. This has lead to the conviction 
that the lepton asymmetry is independent of the low energy 
parameters \cite{lowe}. However, the theoretical input of
the left--right symmetry allows us to relate the right--handed mixing
to the left--handed one and connects the baryon asymmetry to the 
parameters of the left--handed neutrinos.\\ 

If $m_\nu$ is given by Eq.\ (\ref{eq:convmnu}) then knowing 
the neutrino masses and mixing 
angles from oscillation experiments does not help in determining $m_D$ 
because the right hand side in Eq.\ (\ref{eq:convmnu}) 
is quadratic in $m_D$. Given a specific model for $\tilde{m}_D$ and/or 
$M_R$, one can always invert Eq.\ (\ref{eq:convmnu}) to obtain the asymmetry
$\varepsilon$ as was done in e.g.\ 
\cite{lola,laza1,others,kang,goldberg,jean,nezri,faltra1,faltra2,nielsen}. 
The left--right symmetry provides
instead a more natural framework to obtain $\varepsilon$. 
In this case the unitary matrices diagonalizing $m_\nu$ and $M_R$ 
are related. Furthermore, in an interesting situation the 
oscillation experiments provide us with $m_\nu$, which is 
used for the derivation of $M_R$, as described in the next section.

\section{\label{sec:lrmodels}Left--right symmetric models}
The minimal left--right symmetric model \cite{LR} implementing the 
see--saw mechanism requires three Higgs fields, namely: 
a bi--doublet and a right--handed as well as a left--handed 
triplet\footnote{There are various models with triplets of Higgs 
fields \cite{triplet}.}. 
The presence of the latter is necessary in order to maintain the 
left--right symmetry. Both the triplets acquire vevs $v_L$ and 
$v_R$, respectively, at the minimum of the potential. 
Each of them generates a Majorana mass term  
for left-- and right--handed neutrinos: 
\be \label{eq:mlmr}
\bad
m_L = f \, v_L & \mbox{ and } & M_R = f \, v_R
\ea
\ee
with $f$ being the coupling matrix in generation space. The conventional 
see--saw formula (\ref{eq:convmnu}) is then modified to \cite{LR,moh} 
\be \label{eq:typeII}
m_\nu = m_L - \tilde{m}_D \, M_R^{-1} \, \tilde{m}_D^T . 
\ee
Frequently, the first term is neglected which however might not be justified, 
as we will argue below. In fact, whenever $m_L$ is the dominant 
contribution to $m_\nu$ we will have 
\be \label{eq:fapprox}
f = \frac{1}{v_L} \, U_L^T \, {\rm diag}(m_1,m_2,m_3) \, U_L . 
\ee 
Here $U_L$ is the matrix diagonalizing the neutrino mass matrix
$m_\nu$:
\be
\label{eq:uldef}
U_L^\ast \, m_\nu \, U_L^{\dagger} = {\rm diag} (m_1,m_2,m_3)
\ee
and $m_i$ are the light neutrino masses. 
We must however be careful not to ignore the second term in 
Eq.\ (\ref{eq:typeII}) in cases when it is important. 
Later on, we argue that 
when one identifies the Dirac mass matrix with the up quark mass matrix 
then only the top quark gives a sizable contribution.\\

At the minimum of the potential, the left-- and right--handed triplets 
assume their vevs and produce masses for the gauge bosons. Then, in 
general \cite{LR} the following relation holds:  
\be \label{eq:vlvr}
v_L \, v_R \simeq \gamma \, M_W^2 , 
\ee
where the constant $\gamma$ is a model dependent parameter of 
$\cal{O}$(1). Substituting the results from Eqs.\ (\ref{eq:mlmr}) and 
(\ref{eq:vlvr}) into (\ref{eq:typeII}) yields
\be \label{eq:mnulr}
m_\nu = v_L \, 
\left( f - \tilde{m}_D \, 
\frac{f^{-1}}{\gamma M_W^2} \, \tilde{m}_D^T \right) . 
\ee
This result exhibits the strength of the left--right symmetric theory. 
The oscillation experiments allow one to estimate several matrix 
elements of $f$ through Eqs.\ (\ref{eq:typeII}) and (\ref{eq:fapprox}). 
Once we identify $\tilde{m}_D$ with the up quark matrix and 
decide that the contribution from the top quark alone is 
important, we can determine $f$, whose diagonalization gives $U_R$, which 
in turn gives $m_D$ and therefore the lepton asymmetry via 
Eq.\ (\ref{eq:eps}).\\ 

In the next section we will shortly discuss the current status of the 
neutrino mass matrix and will then 
take up the task of calculating $f$, 
estimate the magnitude of $v_{L,R}$ and determine the baryon 
asymmetry within the situation described before.

\section{\label{sec:mnu}Current status of $m_\nu$}
The experimental data on neutrino oscillations can be used to derive 
the neutrino mass matrix \cite{mnu}. The mixing matrix $U_L$ may be 
parameterized as  
\bea \label{eq:Upara}
U_L = U_{\rm CKM} \; 
{\rm diag}(1, e^{i \alpha}, e^{i (\beta + \delta)}) \\[0.3cm]
= \left( \bad 
c_1 c_3 & s_1 c_3 & s_3 e^{-i \delta} \\[0.2cm] 
-s_1 c_2 - c_1 s_2 s_3 e^{i \delta} 
& c_1 c_2 - s_1 s_2 s_3 e^{i \delta} 
& s_2 c_3 \\[0.2cm] 
s_1 s_2 - c_1 c_2 s_3 e^{i \delta} & 
- c_1 s_2 - s_1 c_2 s_3 e^{i \delta} 
& c_2 c_3\\ 
               \ea   \right) 
 {\rm diag}(1, e^{i \alpha}, e^{i (\beta + \delta)}) , 
\eea
where $c_i = \cos\theta_i$ and $s_i = \sin\theta_i$. The ``CKM--phase'' 
$\delta$ may be probed in oscillation experiments, as long as 
the LMA solution is the solution to solar oscillations \cite{osccp}. 
The other 
two ``Majorana phases'' $\alpha$ and $\beta$ can be investigated in 
\onbb{} \cite{Majpha,ichNPB}.  
The choice of the 
parameterization in Eq.\ (\ref{eq:Upara}) reflects this fact since 
the $ee$ element of the mass matrix $\sum_i U_{Lei}^2 m_i$ 
is only depending on the phases 
$\alpha$ and $\beta$. In a hierarchical scheme, to which we will 
limit ourselves, there is no constraint 
on the phases from \onbb{} \cite{ichNPB}. 
Thus, we can choose them arbitrarily. The mass eigenstates are given as 
\be \label{eq:m3m2m1}
\ba
m_3 \simeq \sqrt{\dma + m_2^2} \\[0.4cm]
m_2 \simeq \sqrt{\dms + m_1^2} \gg m_1 . 
\ea
\ee
The values of $\theta_2$ and \dma{} are known to a good 
precision, corresponding to maximal mixing $\theta_2 \simeq \pi/4$ and 
$\dma \simeq 3 \times 10^{-3}$ eV$^2$. Regarding $\theta_1$ and \dms{} 
three distinct areas in the parameter space are allowed, small (large) 
mixing, denoted SMA (LMA) and quasi--vacuum oscillations (QVO): 
\be \label{eq:solsol}
\bad
\mbox{ SMA: } & \tan^2 \theta_1 \simeq 10^{-4} \ldots 10^{-3} , & 
\dms \simeq 10^{-6} \ldots 10^{-5} \, {\rm eV^2} \\[0.4cm]
\mbox{ LMA: } & \tan^2 \theta_1 \simeq 0.1 \ldots 4 , & 
\dms \simeq 10^{-5} \ldots 10^{-3} \, {\rm eV^2} \\[0.4cm]
\mbox{ QVO: } & \tan^2 \theta_1 \simeq 0.2 \ldots 4 , & 
\dms \simeq 10^{-10} \ldots 10^{-7} \, {\rm eV^2} 
\ea . 
\ee
For the last angle $\theta_3$ there exists only a limit of about 
$\sin^2 \theta_3 \ls 0.08$.
For a recent three--flavor fit to all available data see \cite{carlos}.\\ 

Note that we have identified the neutrino mixing matrix 
in Eq.\ (\ref{eq:Upara}) with the matrix $U_L$ diagonalizing the neutrino
mass matrix Eq.\ (\ref{eq:uldef}). 
This assumes implicitly that the charged lepton mixing
is small. We shall work with this assumption in what follows.

\section{\label{sec:detf}Determination of $f$ and the baryon asymmetry}

As mentioned before, we argue that only the top quark gives a sizable 
contribution to the conventional see--saw formula 
$\tilde{m}_D \, M_R^{-1} \, \tilde{m}_D^T $. 
Identifying $\tilde{m}_D$ with the up quark mass matrix 
and neglecting mixing among up quarks, the relative magnitude of both
terms contributing to $m_\nu$ 
can be written as 
\be \label{eq:estlr}
\frac{\D |\tilde{m}_D \, M_R^{-1} \, \tilde{m}_D^T |}{\D |m_L| } \simeq 
\frac{m_q^2/v_R}{\D v_L} \simeq  
\frac{\D m_q^2}{\D \gamma \, M_W^2} , 
\ee
where we only used Eq.\ (\ref{eq:vlvr}) and assumed that the 
matrix elements of $f$ and $f^{-1}$ are of the same order of magnitude.  
One sees immediately that only the top quark mass makes the ratio 
in Eq.\ (\ref{eq:estlr}) of order one. 
In practically all models \cite{models} the heaviest mass is the 
(33) entry of the mass matrix, which means that only the 
(33) element of $m_\nu$ has a contribution from the term 
$\tilde{m}_D \, M_R^{-1} \, \tilde{m}_D^T $. The matrix $\tilde{m}_D$ 
may therefore be taken as 
\be \label{eq:tmd}
\tilde{m}_D \simeq {\rm diag} (0,0,m_t)   .  
\ee
There might be a common factor of order 1 for the complete matrix, but 
in light of the factor $\gamma$ in Eq.\ (\ref{eq:vlvr}) and the 
uncertainty in the oscillation parameters we can safely work with 
this form of $\tilde{m}_D$. Later on we will comment on the dependence 
of the results on this factor.\\
 
It is helpful to repeat the 
argument with typical numbers. The maximal scale of $m_\nu$ is 
$\sqrt{\dma} \simeq 0.1$ eV\@. Then the relations 
$v_L \, v_R = \gamma \, M_W^2$ and 
$m_\nu \simeq f \, v_L \simeq 0.1 $ eV are compatible for 
$v_R \simeq 10^{15}$ GeV and thus $v_L \simeq 0.1$ eV as long as 
$f \simeq 0.1 \ldots 1$. The scale of $v_L \, f$ is again matched 
by the factor $m_t^2/v_R$. This means that $v_R$ is close to the 
grand unification scale and $v_L$ is of order of the neutrino masses, 
which one expects since $m_L$ is the dominating contribution to 
$m_\nu$.\\

We can now proceed to calculate the contribution to the Yukawa 
coupling matrix $f$ in this situation. Since only the (33) element 
of $m_\nu$ has a contribution from the see--saw term we have
\be \label{eq:fe33}
f_{ij} = \frac{(m_\nu)_{ij}}{v_L} \; \mbox{ for all } i,j \; 
\mbox{ except for } i = j =3 .
\ee
For the last term we adopt 
\be \label{eq:f33}
f_{33} = \frac{(m_\nu)_{33} + s}{v_L} , 
\ee
where the parameter $s$ denotes the contribution arising from the 
see--saw term. The parameter is consistently determined 
by using Eqs.\ (\ref{eq:vlvr},\ref{eq:mnulr}) 
\be
s = \left( \frac{\tilde{m}_D \, f^{-1}\, \tilde{m}_D^T}{v_R} 
\right)_{33} = \frac{m_t^2}{v_R} \frac{F_{33}}{\rm det \, f} = 
\frac{m_t^2}{v_R} \frac{F_{33}}{\tilde{F} + F_{33} \, f_{33}}
\ee
where $F_{33} = f_{11} \, f_{22} - f_{12}^2$ and 
$\tilde{F} = 2\, f_{12} \, f_{13} \, f_{23} - f_{13}^2 \, f_{22} - 
f_{23}^2 \, f_{11}$. Using Eq.\ (\ref{eq:f33}) we can solve for $s$ 
and find 
\be \label{eq:s}
s \simeq \pm \frac{\sqrt{\gamma} \, M_W \, m_t}{v_R} . 
\ee 
As expected, $s$ is of the order of 0.1 to 0.01 eV\@.\\
 
With the matrix $f$ now determined completely, we diagonalize it and 
calculate the baryon asymmetry in the following way.  
From $\varepsilon$ the baryon asymmetry $Y_B$ is obtained by 
\be \label{eq:YBth} 
Y_B = c\, \kappa \frac{\varepsilon}{g^\ast} , 
\ee
where $g^\ast \simeq 110 $ is the effective number of massless degrees of 
freedom at $T = M_1$. The factor $c$ indicates the fraction of the 
lepton asymmetry converted into a baryon asymmetry via the sphaleron 
processes \cite{sphaleron}. 
It depends on the group structure of the theory and is of order one. 
For three lepton families and one (also two) Higgs doublets it 
is approximately equal to $-0.55$. 
Finally, $\kappa$ is a dilution factor due to 
lepton--number violating wash--out processes. 
It can be obtained by integrating 
the Boltzmann equations and depends strongly on 
\be \label{eq:K}
K \equiv \frac{\Gamma_1}{H(T = M_1)} = 
\frac{ (m_D^\dagger m_D)_{11} \, M_1}{8\, \pi \, v^2}  
\frac{M_{\rm Pl}}{ 1.66 \, \sqrt{g^\ast} \, M_1^2} , 
\ee 
where $\Gamma_1$ is the width of the lightest Majorana neutrino and 
$H(T = M_1)$ the Hubble constant at the temperature of the decay. 
$M_{\rm Pl}$ is the Planck mass.  
A convenient parameterization is \cite{param}
\be \label{eq:kappa}
-\kappa \simeq 
\left\{ 
\bad 
\sqrt{0.1 \, K} \, \exp (-4/3 \, (0.1\, K)^{0.25}) 
& \mbox{ for } & K \gs 10^6 \\[0.4cm]
\frac{\D 0.3}{\D K \, (\ln K)^{0.6} }
&  \mbox{ for } & 10 \ls K \ls 10^6 \\[0.4cm]
\frac{\D 1}{\D 2 \, \sqrt{K^2 \, + 9}} 
& \mbox{ for } & 0 \ls K \ls 10
\ea 
\right. . 
\ee
Typically, values for $\kappa$ lie in the range of $10^{-3}$ to 0.1.\\

In addition to the decay of the right handed neutrino, the 
out--of--equilibrium decay of the Higgs triplet has also been considered as
possible mechanism for generating lepton asymmetry \cite{ma}. This
needs CP violation in the Higgs sector and hence an 
enriched Higgs sector to implement it. 
For example, the models in \cite{ma,lazarides,chun}
need the presence of two left--handed Higgs triplets as opposed to one
triplet considered here. 
If an asymmetry is produced with several higgs triplets, their
subsequent decays it will tend to be erased as long as the triplets are
heavier than the lightest right--handed neutrino. 
This mass pattern happens to be the natural possibility 
in the present scenario.  The mass of the Higgs triplet
is given by $\lambda v_R$ where $\lambda$ is a typical quartic coupling
of  the Higgs potential. In contrast, the mass of the lightest 
right--handed neutrino
is given within our approximation \cite{eapjos} by $M_1 \sim \alpha v_R$,
where $\alpha \equiv{m_2 s_1^2\over m_3}$ lies in the range
 $(10^{-3}-10^{-6})$, depending
upon the chosen solution for the solar neutrino problem. Hence, the
triplet will be heavier than the lightest right--handed 
neutrino as long as the quartic coupling $\lambda$ is 
$\cal O$(1). The lepton asymmetry created through
triplet will be washed out in this case, according to 
the usual damping of any preexisting asymmetry. For the above reasons 
the asymmetry originating from the triplet decay does not contribute 
to the following numerical analysis.

\section{\label{sec:res}Results}

The main variables are the parameters \dms{} and $\tan^2 \theta_1$ 
which specify the solar solution, as 
given in Eq.\ (\ref{eq:solsol}). 
Below, we analyze the dependence of $Y_B$
on these parameters. It is found that the value of $\gamma$ and 
the sign of $s$ do not play a decisive role. Also, the value of 
\dma{} (varied within $(3 \pm 5) \cdot 10^{-3}$ eV$^2$)  
has little influence on $Y_B$. The same is true for changing 
$\tan^2 \theta_2$. 
The asymmetry decreases (increases) 
with decreasing (increasing) top quark mass, though not much. 
For the SMA case the dependence on the phase $\alpha$ is 
not as strong as on the 
other two phases, whereas it is equally strong for the LMA and QVO case. 
The conclusions we draw now will be only changed 
if all these parameters conspire and take rather extreme values within their 
allowed ranges.\\ 

\begin{figure}[hp]
\epsfig{file=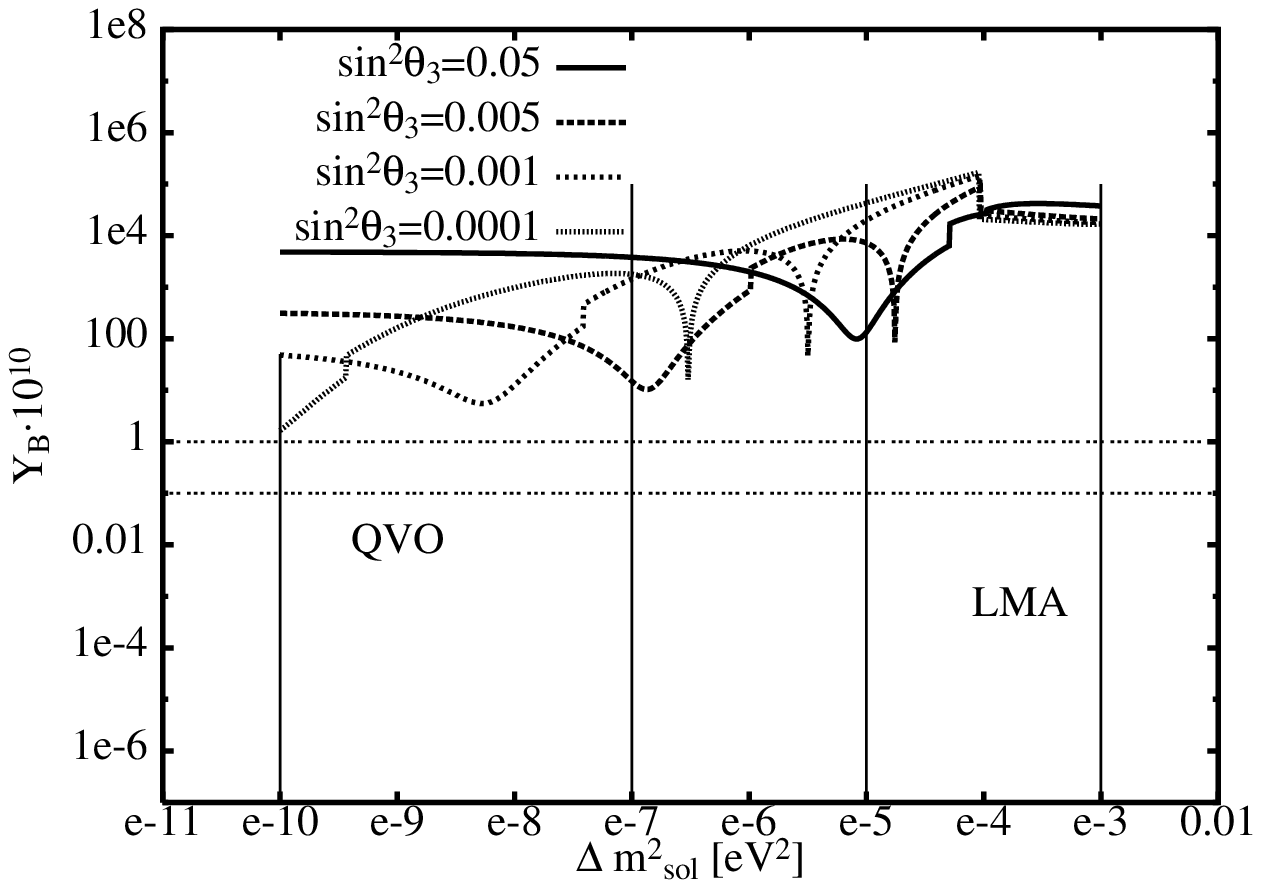,width=13cm,height=8cm}
%\vspace{0.5cm}
\caption{\label{fig:LMdm2}Behavior of the baryon asymmetry as a function of 
\dms{} for different $\sin^2 \theta_3$. For this plot we fixed 
$\dma = 3 \cdot 10^{-3}$ eV$^2$, $\theta_1 = \theta_2 = \pi/4$, 
$\alpha = \pi/3$, $\beta = \pi/5$ and $\delta = \pi/6$.}
\vspace{0.5cm}
\epsfig{file=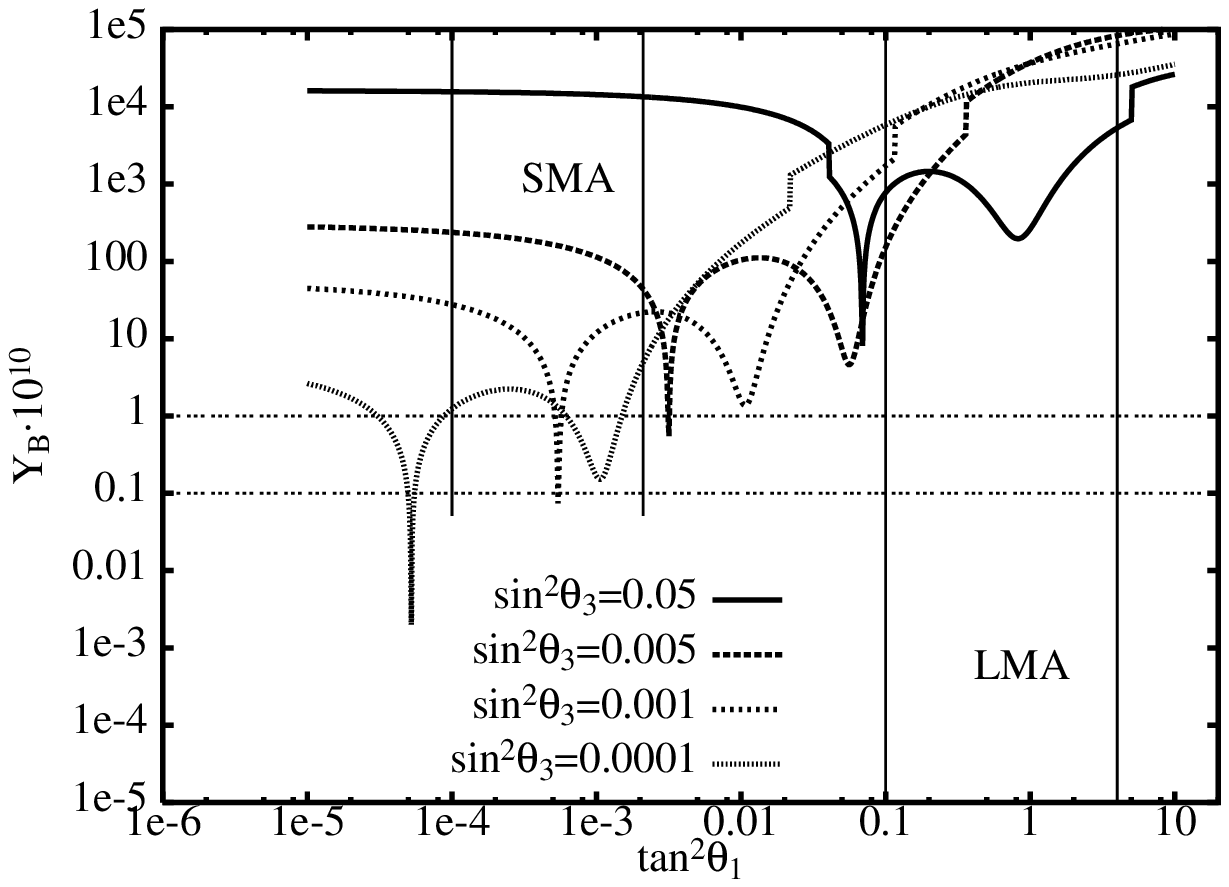,width=13cm,height=8cm}
%\vspace{0.5cm}
\caption{\label{fig:t12}Behavior of the baryon asymmetry for 
$\dms = 10^{-5}$ eV$^2$ 
as a function of 
$\tan^2 \theta_1$ for different $\sin^2 \theta_3$. The phases are 
$\alpha = \pi/3$, $\beta = \pi/4$, $\delta = \pi/6$ and 
the other parameters 
are as in the previous plot.}
\end{figure}

We work now with positive 
$s$ from Eq.\ (\ref{eq:s}) and apply maximal atmospheric mixing 
with $\dma = 3 \cdot 10^{-3}$ eV$^2$. 
The parameter $\gamma$ is fixed to one and the top quark mass 
at 175 GeV\@. We shall work with $v_R = 10^{15}$ GeV 
from which $v_L$ and $s$ are obtained via 
Eqs.\ (\ref{eq:vlvr}) and (\ref{eq:s}). 
We find that $K$ from Eq.\ (\ref{eq:K}) 
is always below 10 and thus $\kappa$ lies between $0.17$ and $0.05$. 
There is practically no dependence 
on the lightest mass eigenstate $m_1$.  
Fig.\ \ref{fig:LMdm2} shows the behavior 
of $Y_B$ as a function of \dms{} for different $\sin^2 \theta_3$. One sees 
that for lower masses the asymmetry decreases. 
In Fig.\ \ref{fig:t12} we display the dependence on $\tan^2 \theta_1$. 
The baryon asymmetry decreases with decreasing $\tan^2 \theta_1$. 
This dependence is stronger than the one on \dms. 
In both plots it is seen that $Y_B$ approximately increases with 
increasing $\sin^2 \theta_3$.\\
 
\begin{figure}[hp]
\epsfig{file=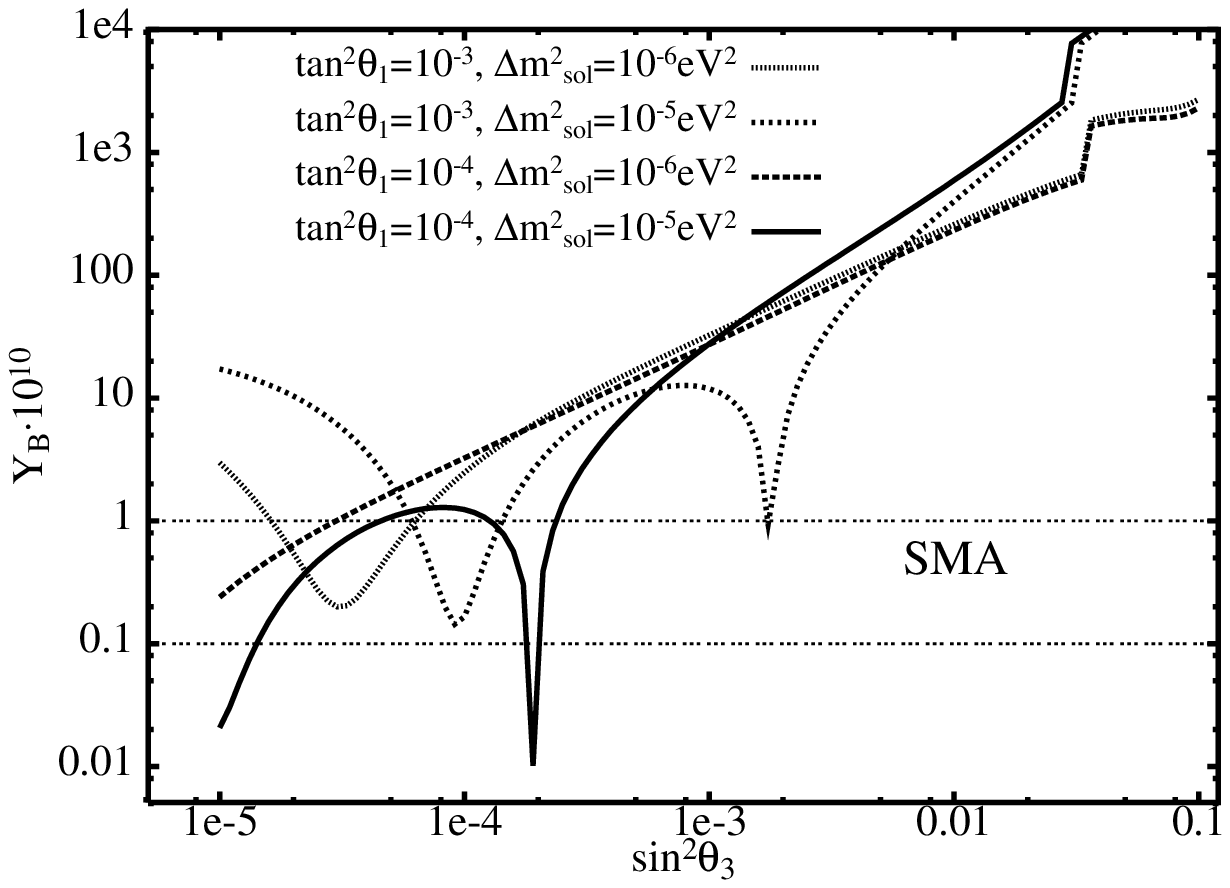,width=13cm,height=8cm}
%\vspace{0.5cm}
\caption{\label{fig:SMA}Behavior of the baryon asymmetry as a function of 
$\sin^2 \theta_3$ for different \dms{} and $\tan^2 \theta_1$ for the 
case of the SMA solution. 
The other parameters are as in the previous plot.}
\vspace{0.5cm}
\epsfig{file=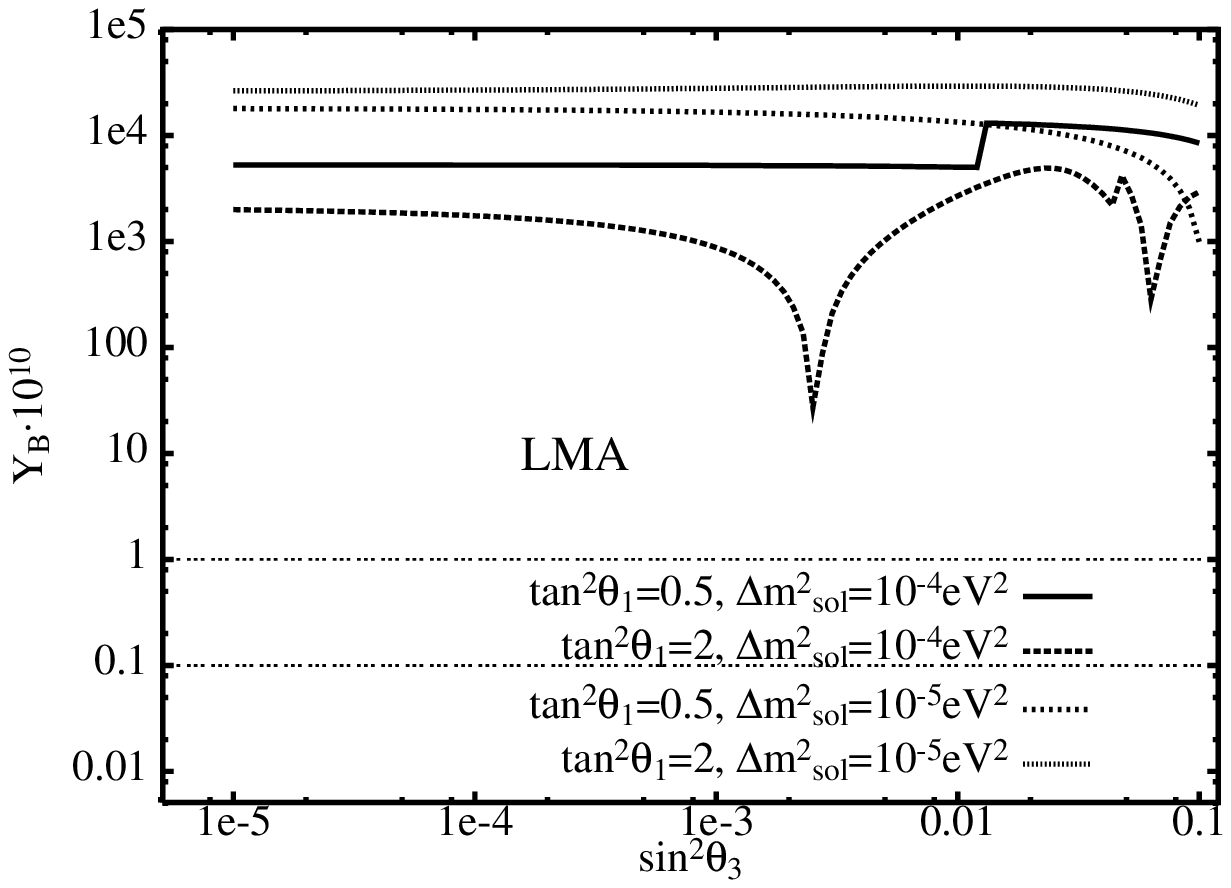,width=13cm,height=8cm}
%\vspace{0.5cm}
\caption{\label{fig:LMA}Behavior of the baryon asymmetry as a function of 
$\sin^2 \theta_3$ for different \dms{} and $\tan^2 \theta_1$ for the 
case of the LMA solution. 
For this plot we fixed the atmospheric parameters as before and 
$\alpha =\pi/5$, $\beta = \pi/6$ and $\delta = \pi/3$.}
\end{figure}

\begin{figure}[hp]
\epsfig{file=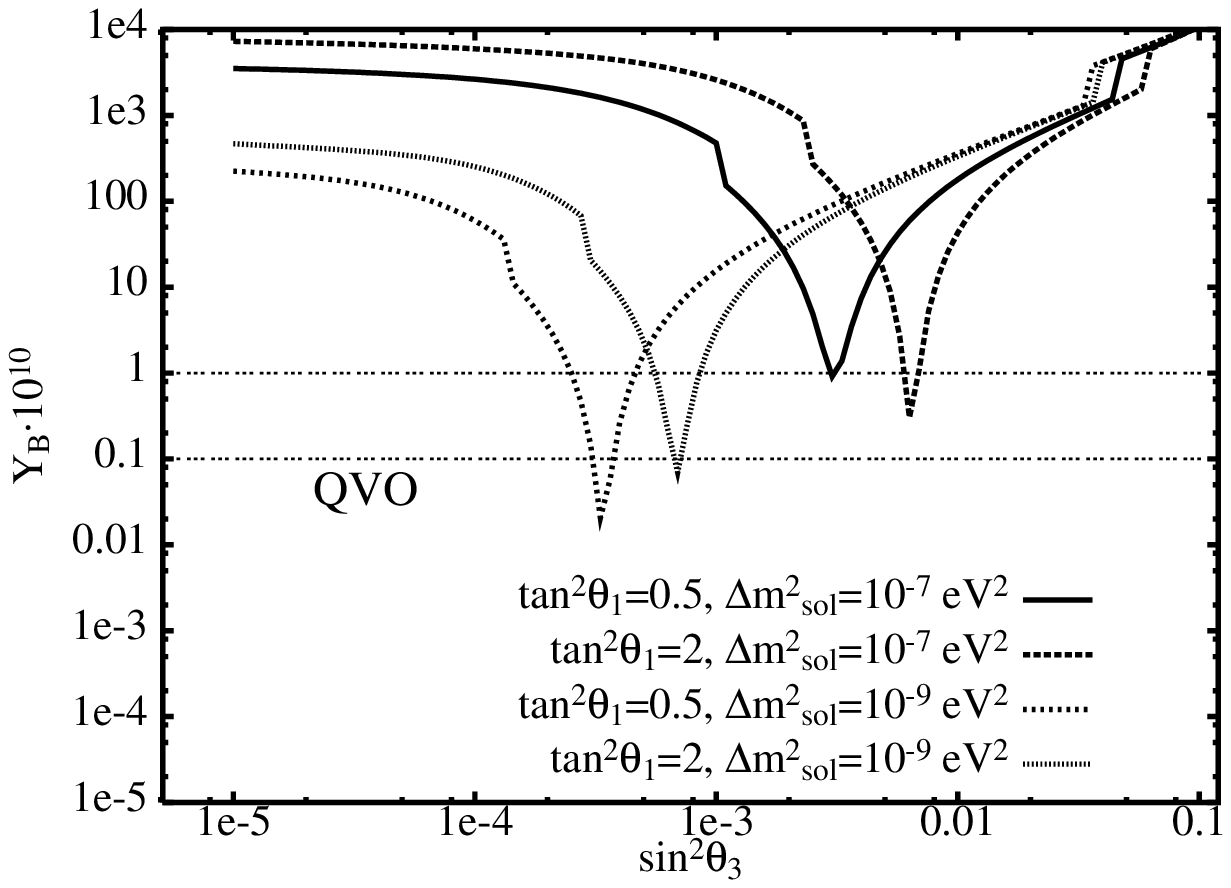,width=13cm,height=8cm}
\vspace{0.5cm}
\caption{\label{fig:QVO}Behavior of the baryon asymmetry as a function of 
$\sin^2 \theta_3$ for different \dms{} and $\tan^2 \theta_1$ for the 
case of the QVO solution. 
For this plot we fixed the atmospheric parameters as before and 
$\alpha = \pi/5$, $\beta = \pi/4$ and $\delta = \pi/6$.}
\end{figure}

We analyze next the three distinct solutions to the solar neutrino 
problem in detail. Fig.\ \ref{fig:SMA} shows the SMA case for different 
values of the parameters. All four combinations yield $Y_B$ in the right 
magnitude and seem to prefer a $\sin^2 \theta_3$ lower than about $10^{-3}$. 
Fig.\ \ref{fig:LMA} shows again that the  
LMA case results in a very high asymmetry. Here, 
fine--tuning of the parameters, specifically the $CP$ violating phases is
required to get a $Y_B$ within its 
experimental limits. It is also seen that $\tan^2 \theta_1 > 1$ gives a 
smaller asymmetry than $\tan^2 \theta_1 < 1$.  
The QVO case, displayed in 
Fig.\ \ref{fig:QVO}, might also produce an acceptable asymmetry. 
Note the different choice of the phases in this plot 
and Fig.\ \ref{fig:LMdm2}. We note that the latest SuperKamiokande data 
seems to favor the LMA solution \cite{SK}, using however a 
two--flavor analysis. For a more definitive conclusion additional 
data has to be waited for.\\ 

All our results are based on identifying $\tilde{m}_D$ with the up quark
mass matrix and retaining only the top quark contribution. The importance
of the ordinary see--saw contribution will be less in any other models in
which the largest scale of $\tilde{m}_D$ is set by a fermion mass other than
the top quark, i.e. the bottom quark mass. In the extreme case of
completely neglecting the ordinary see--saw contribution, one will  
have $U_R = U_L$ and $M_{Ri} = \frac{v_R}{v_L} m_i$. In this case, 
the lepton asymmetry will be completely controlled by the 
left--handed neutrino masses and mixings as well as the ratio 
$\frac{v_R}{v_L}$. Thus, unlike in the present
case, the results will be sensitive to the value of $m_1$ which has to be
less than or similar to the solar scale. We have checked that the required
asymmetry can be generated in this case with a proper choice of $m_1$.\\
 
It is instructive to see the dependence of some other models on 
the solar solution. Models based on $SO(10)$ were used e.g.\ in 
\cite{nezri} where it was found that only the QVO solution gives acceptable 
baryon asymmetry. A  slightly different analysis in \cite{faltra1} finds that 
also the SMA case gives an  acceptable asymmetry. 
This solution has also been favored in the models 
presented in \cite{kang,goldberg,nielsen}, which all use quite different 
symmetries. The LMA 
solution, which we disfavor, has been shown to be the 
only solution producing an acceptable $Y_B$ 
in \cite{faltra2}, using $SU(5)$ inspired mass matrices. We stress again 
that the main difference to the present paper lies in the fact that 
the left--handed Higgs triplet plays a dominant role in producing 
the light neutrino mass matrix. 
Once one solution for the solar oscillation 
is established, more definite statements about the symmetry relations 
can be made, which is an important by--product of the 
analysis of relations between leptogenesis and neutrino oscillations.

\section{\label{sec:concl}Conclusion}
Using very general properties of left--right symmetric theories 
we connected the light left--handed neutrino 
sector as measured in neutrino oscillations with heavy right--handed 
neutrinos, whose decay is responsible for the baryon asymmetry of the 
universe. Identifying the Dirac mass matrix with the up quark mass 
matrix we found that only the top quark has a significant contribution to the 
neutrino mass matrix. 
The main contribution to $m_\nu$ comes from the left--handed triplet, which 
is neglected in most papers dealing with this subject. 
In our scenario, the SMA and QVO case yield in reasonable asymmetry, 
whereas the LMA solution produces an asymmetry which is too high.

\hspace{3cm}
\begin{center}
{\bf \large Acknowledgments}
\end{center}
This work has been supported in part by the
``Bundesministerium f\"ur Bildung, Wissenschaft, Forschung und
Technologie'', Bonn under contract No. 05HT1PEA9.
Financial support from the Graduate College
``Erzeugung und Zerf$\ddot{\rm a}$lle von Elementarteilchen''
at Dortmund university is gratefully acknowledged (W.R.). W.R.\ wishes 
to thank the Universita degli studi di Pisa 
where the final stages of this work were performed. A.S.J.\ acknowledges
support from the Alexander von Humboldt foundation and 
thanks his colleagues for their 
hospitality at the University of Dortmund.

\end{document}